\documentclass[twocolumn]{phyeauth}

 \usepackage{graphics}
\usepackage{graphicx}
\usepackage{epsfig}

\begin{document}

\begin{frontmatter}

\title{Phenomenology of the 0.7 conductance feature}

\author{D. J. Reilly\corauthref{cor1},}
\author{ Y. Zhang and}
\author{L. DiCarlo}
\corauth[cor1]{Email: reilly@fas.harvard.edu}
\address{Department of Physics, Harvard University, Cambridge, 02138, USA}
\begin{abstract}
We describe a phenomenological model for the conductance feature near $0.7 \times 2e^2/h$ that occurs in quantum point contacts. We focus on the transconductance at finite source-drain bias and contrast our model with the results expected from a single-particle picture. Good agreement is seen in comparing the model with experimental data, taken on ultra-low-disorder
GaAs induced electron systems. Although simple, our phenomenology suggests 
important boundary conditions for an underlying microscopic theory. 
\end{abstract}

\end{frontmatter}

\section{Introduction}
The conductance of a quantum point contact (QPC), formed via geometric or electrostatic confinement of a two-dimensional electron system (2DES), is quantized in units of $2e^2/h$ \cite{vanWees_QPC,Wharam_QPC1st}.  This quintessentially 
quantum-mechanical phenomena is well explained in terms of the single-particle discrete energy spectrum associated with the confining potential \cite{Buttiker_QPC}. QPCs also exhibit unexpected  
non-quantized features that cannot be explained within this non-interacting, single-particle  picture.  
A clear example is the feature occurring below the first conductance plateau, in the range 
$0.5 - 0.7 \times 2e^2/h$ and at equivalent positions for higher plateaus. This anomalous conductance structure was observed in the earliest transport measurements on ballistic QPCs \cite{vanWees_QPC,Wharam_QPC1st} but 
has remained largely unexplained despite numerous experimental  investigations 
\cite{Thomas_spin1st,Thomas_int,BK_QWAPL,Kristensen,KristensenPRB,Nuttinck,newthomas,ctliang,Pyshkin,Reilly_PRB01,Appleyard,Cronenwett_07,Reilly_PRL02,bird,Roche_fano,depicciotto}  and is regarded as a key outstanding problem in mesoscopic physics \cite{Fitzgerald}. 

A prime result, initially uncovered by Thomas {\it et al.,} \cite{Thomas_spin1st} is the evolution of the
 $0.7 feature$ with increasing parallel magnetic field into the well understood Zeeman spin-split plateau at $0.5 \times 2e^2/h$. This  observation strongly suggests that the $0.7$ structure at $B=0$ is related to a  many-body state in which symmetry is broken between spin-up and spin-down electrons.  
\section{Phenomenological model}
Here we describe a simple phenomenological model that is in striking 
agreement with the experimental behavior of the conductance feature with temperature $T$, magnetic field $B$, source-drain bias $V_{sd}$ and potential profile.
At the core of this model is the conjecture that a density-dependent spin-gap 
opens in the one-dimensional (1D) energy spectrum as the gate voltage is made 
less negative \cite{Reilly_PRB01,Reilly_PRB05}. 
Fig. 1a captures the essence of the model. Before the first 1D sub-band is occupied 
(below energy $E1$ in Fig. 1a) the spin-up and 
spin-down sub-bands are degenerate. As the 1D states begin to fill 
the spin-down band rapidly moves down in energy with
the available spin-up states moving higher in energy above the Fermi-level ($E_F$).
The model assumes a parabolic dependence of the Fermi-energy on gate voltage ($E_F \sim V_g^2$) in connection with the usual $E^{-1/2}$ form of the 1D density of states \cite{Wingreen1}. For a detailed discussion of this model we refer the reader to Refs. \cite{Reilly_PRB05,Reilly_PRL02}.

\begin{figure}[t!]
\includegraphics[width=7.5cm]{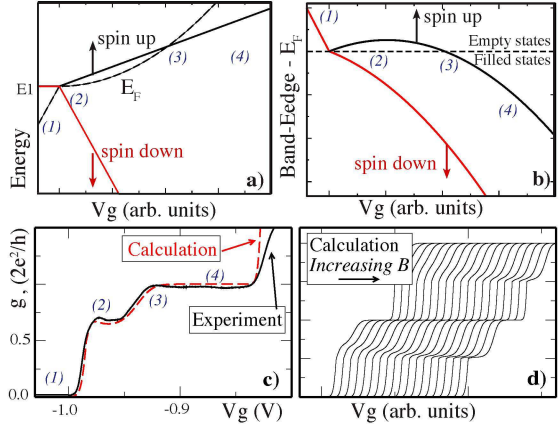}
\caption{a) Energy level configuration underlying our phenomenological model. A spin-gap opens in the 1D energy spectrum as a function of gate bias $V_{g}$. The Fermi-level $E_{F}$ is parabolic with gate voltage because of the singularity in the 1D density of states. b) Shows the {\it difference} between 
$E_F$ and the spin-up and spin-down band-edges. c) Shows conductance calculations based on our model (red and dashed line) together with experimental data (black and solid line) taken at $T$=50mK. d) Shows calculated conductance in the context of the model for increasing parallel magnetic field $B$. Consistent with experiment we observe a smooth evolution from $0.7$ to $0.5 \times 2e^{2}/h$. Panels c) and d) are taken from Reference \cite{Reilly_PRB05}. } 
\end{figure}

Features in conductance are related to the occupation probability of the 1D sub-bands and are a function  of temperature and the energy difference between 
the sub-band edge and $E_F$. Within the context of our model, Fig. 1b shows 
the difference between the spin-dependent band-edge energy and $E_F$ as a function of gate voltage $V_g$.    
It is the functional form of this {\it difference} between the spin band-edges and $E_F$ that relates 
to the experimental data (since it sets the conductance) and of prime importance in suggesting possible 
`boundary conditions' of an underlying microscopic theory. We point out that in contrast to a similar model based on pinning of the band-edge \cite{Kristensen_Bruus}, in our picture the spin-down energy continues to move rapidly below the Fermi-level as the gate voltage is swept.

Figure 1c (taken from Ref \cite{Reilly_PRB05}) compares differential conductance $g=di/dv$ calculations based on this model with data taken on an ultra-low-disorder quantum wire at $T$=50mK 
(See Refs \cite{Reilly_PRB01,Reilly_PRL02,Reilly_PRB05} for experimental details). We take the simplest approximation and calculate conductance with a step transmission function in the limit of zero source-drain bias. Importantly, the only 
free-parameter needed to fit the model to the experimental data is the 
rate at which the spin-gap opens with gate voltage $\gamma = dE_{\uparrow\downarrow}/dV_g$ \cite{Reilly_PRB05}.
Note the small discrepancy between model and experiment at the top right of Fig. 1c which 
is due to the step transmission 
function (no tunneling) that is used in the calculation rather than a more realistic smoothly 
varying $T(E)$ \cite{Buttiker_QPC}. Fig. 1d 
(also taken from Ref. \cite{Reilly_PRB05}) shows conductance calculations as a function of increasing 
parallel magnetic field, $B_{\parallel}$ in which a smooth evolution of the conductance feature from $0.7$ to $0.5 \times 2e^2/h$ is observed, consistent with the experimental results \cite{Thomas_spin1st,Cronenwett_07,Graham}.

\section{Single-particle results}   
We now focus on the transconductance ($dg/dV_g$) at finite $V_{sd}$ which 
facilitates the study of transitions between the conductance 
plateaus as a function of chemical potential or gate voltage $V_g$. We firstly 
compare transconductance plots based on our model to the single-particle case at 
zero and finite magnetic field. 
Fig. 2a shows a transconductance intensity plot as a function of $V_{sd}$ and $V_g$ 
calculated using the usual single-particle formalism \cite{Glazman3,moreno}. 
We assume the conductance at finite bias is approximated by the weighted average of two zero $V_{sd}$ conductances, one for the chemical potential of the source ($E_F+\beta eV_{sd}$) and the other for the drain ($E_F-(1-\beta)eV_{sd}$), where $\beta$ characterizes the symmetry of the potential drop across the constriction \cite{moreno}. Centered 
about $V_{sd}$=0 are the 
linear response integer-plateau diamonds (dark) 
at $0,1$ and $2 \times 2e^2/h$. Symmetric either side of the integer 
plateaus are the finite bias `half-plateaus'
that occur when the chemical potential of the source ($\mu_s$) or drain ($\mu_d$) differ 
by one (spin degenerate) 
sub-band ($g=0.5$ and $1.5 \times 2e^2/h$) \cite{Patel_QPCBpar}. 

\begin{figure}[t!]
\includegraphics[width=7.5cm]{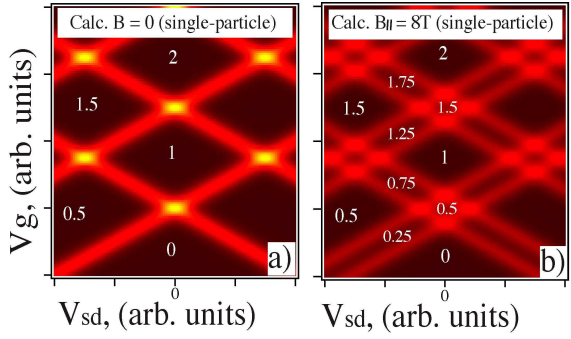}
\caption{Single-particle (non-interacting) transconductance $dg/dV_g$  calculations for $B$=0 (a) and $B_\parallel$=8T (b). Transconductance is plotted on the  intensity-axis in arbitrary units as a function of souce-drain bias $V_{sd}$ and gate voltage $V_g$. Following Glazman and Khaetskii \cite{Glazman3} we observe integer plateau diamonds (dark areas) centered at $V_{sd}$=0 (labeled  0, 1 and 2...). Either side of the integer plateaus are the finite bias half-plateuas (labeled 0.5 and 1.5...) that occur when $\mu_s$ and $\mu_d$ differ by two spin bands (a). In (b) the finite $B$-field Zeeman spin splitting results in zero-bias plateaus (labeled 0, 0.5, 1, 1.5 and 2...) and finite-bias plateaus (labeled 0.25, 0.5, 0.75, 1.25, 1.5, 1.75....).  }
\end{figure}
With the application of an in-plane magnetic field $B_\parallel$ the spin degeneracy is lifted and 
additional Zeeman spin-split 
plateaus appear at $0.5, 1.5 \times 2e^2/h$ at $V_{sd}$=0 as shown in Fig. 2b.
At finite bias and high magnetic field new plateaus also 
appear at $0.25, 0.75, 1.25, 1.75 \times 2e^2/h$. These are the finite bias `half-plateaus' that now occur at quarter intervals with the lifting of the spin 
degeneracy. The relative size of these `quarter-plateaus' in comparison to the half-plateaus simply 
depends on the relative Zeeman splitting to sub-band energy spacing. In Fig. 2b we show 
calculations for $B_{\parallel}$=8T, setting the in-plane $g$-factor to the 
bulk value of 0.44. 
\section{Comparison of model with experiment}
Returning to our phenomenology, Fig. 3a shows transconductance calculations 
based on the model. At low gate voltage the sub-bands are spin-degenerate and 
the transconductance looks similar to the $B$=0 single-particle case of 
Fig. 2a (in the range $g < 0.5 \times 2e^2/h$). With increasing 1D density 
(making the gate voltage less negative) the degeneracy
begins to lift as a spin-gap opens in the 1D energy spectrum. 
At $g > 0.5 \times 2e^2/h$ the transconductance 
now resembles Fig. 2b as the spin degeneracy is lifted and finite-bias plateaus 
beginning at $0.75 \times 2e^2/h$ are observed.

Figure 3b shows data taken on an ultra-low-disorder quantum wire at $T$=100mK. In these devices the 
electrons in both the 1D 
and 2D regions are induced via the application of positive bias to a surface gate \cite{BK_QWAPL}. 
Close examination of the experimental data reveals an absence of finite bias plateaus 
at $0.25 \times 2e^2/h$ and the 
presence of $0.75 \times 2e^2/h$ features, resembling our calculated results in Fig. 3a. 
This behavior is consistent with a picture
of a density-dependent spin gap opening with gate voltage.
\begin{figure}[t!]
\includegraphics[width=7.5cm]{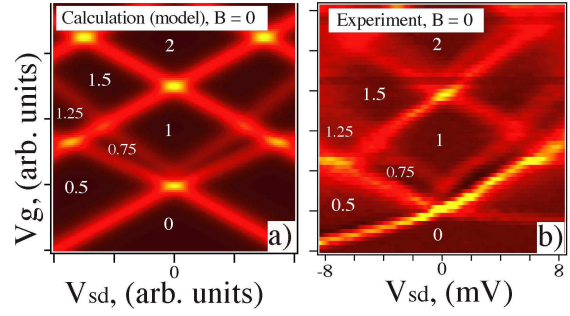}
\caption{Comparison of model and experiment. a) Transconductance $dg/dV_{g}$ calculations based on the phenomenological model, for $B$=0 (transconductance is plotted on the  intensity-axis in arbitrary units as a function of souce-drain bias $V_{sd}$ and gate voltage $V_{g}$). Note the similarity with the finite $B$-field case of Fig. 2b), but with an  absence of low-bias `quarter-plateaus' at $0.25, 1.25... \times 2e^2/h$ consistent with a gate-dependent spin-gap. b) Shows transconductance data for an ultra-low-disorder QPC, formed using induced-electron gating techniques \cite{BK_QWAPL,Reilly_PRB01}  $B$=0, $T$=100mK. Data is the numerical derivative with respect to $V_{g}$ of data in Fig. 5a, Ref. \cite{Reilly_PRB05}.}
\end{figure}
Of particular interest is the occurrence of weak features at $1.25 \times 2e^2/h$, beyond 
the $V_{sd}$ needed to produce the usual half-plateaus. These weak structures, which 
are present in both our calculations and experimental data, connect
to the usual $0.7 \times 2e^2/h$ feature via the line that intersects the top half of 
the first ($2e^2/h$)  diamond (either side of the label `1'). In the context of the model these high $V_{sd}$ $1.25$ features arise from the chemical potential of the source and drain differing by 3 spin sub-bands, as in the case where the source is at $2 \times 2e^2/h$ (above 4 spin band-edges) and the drain is at $0.5 \times 2e^2/h$ (in the gap between the first spin-up and spin-down sub-bands). 
The presence of high bias $1.25$ features provides further evidence that a spin dependent energy gap remains open well below the Fermi-level, continuing  to increase (slightly) as the 1D density grows and higher sub-bands are populated.

\section{0.7 analogs}
We now discuss our phenomenological model in the context of a recent experiment by Graham {\it et al.,} \cite{Graham},   in which the sub-band energies can be made to cross at very high in-plane 
magnetic fields. In addition to the usual evolution of the $0.7$ feature towards $0.5 \times 2e^2/h$, Graham {\it et al.,} discovered the appearance of similar conductance structure at high $B_{\parallel}$ evolving from 
 $1.5 \times 2e^2/h$ into the plateau at $2e^2/h$. This feature is termed a `$0.7$-analog'. In Fig. 4 we show transconductance calculations based on our model as a function of increasing Zeeman energy (bottom to top). For simplicity we ignore any changes in the electron $g$-factor related to correlation or 
diamagnetic effects, taking the bulk value of 0.44. This likely accounts for the larger magnetic fields required in our model to achieve sub-band crossing in comparison to experimental observation.
\begin{figure}[t!]
\includegraphics[width=7.5cm]{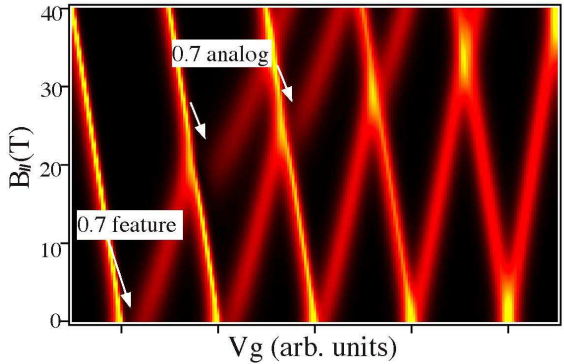}
\caption{Transconductance calculations based on the model for $V_{sd}$=0 as a function of gate bias $V_{g}$ and in-plane magnetic field, $B$ (traces are off-set). We use a bulk value for the electron $g$-factor of 0.44. See main text and Ref. \cite{Graham} for detailed discussion.}
\end{figure}

Consistent with the experimental results of Graham {\it et al.,} \cite{Graham} we observe the usual $0.7$ feature at $B$=0, indicated in Fig. 4 by the black region separating the spin-up and spin-down transconductance lines that evolve away from each other as $B$ is increased. Without adding any new parameters to our model we also find good agreement with the experimental results at magnetic fields where the $N$=1 spin-up sub-band crosses the $N$=2 spin-down sub-band. As in the case of the $0.7$ structure at $B$=0, our model assumes that at high $B_{\parallel}$ as each spin-band is populated, the Fermi energy $E_F$ is proportional to $V_g^2$ and the gate-dependent spin gap (intrinsic) adds to the (extrinsic) Zeeman energy of the $B$-field. 

Following the crossing of sub-bands there is a discontinuous shift in $\delta V_{g}$ from the crossing point, marking the appearance of the $0.7$ analog \cite{Graham}. Our calculations also account for the appearance of  higher order $0.7$ analogs at the crossings of $N$=2 spin-up with $N$=3 spin-down sub-bands. The temperature dependence of the $0.7$ analogs follows the usual activated behavior of the $0.7$ feature in our model (not shown). Very recent numerical work \cite{Berggren_PRB05} using spin-density functional theory also shows good agreement with the experimental results of Graham {\it et al.}. How the functional form of our  phenomenology relates to these more detailed calculations is an interesting question for future work.

The strong agreement between the transconductance calculations and the experimental data of Graham {\it et al.,} provide further evidence that this model captures the essence of the functional form underlying the 0.7 feature. Despite this agreement however, the observation of a zero bias anomaly  in the bias spectroscopy and the low temperature {\it restoration} of the conductance as observed by Cronenwett {\it et al.,} \cite{Cronenwett_07} cannot be explained without extending this model to also include Kondo spin screening in the regime $T<T_K$, where $T_{K}$ is the Kondo temperature. Perhaps a complete explanation will account for a density-dependent gap that is screened by the formation of a Kondo-like state at low temperatures and bias.

\section{Conclusions}
We have described a simple phenomenological model for the $0.7 \times 2e^2/h$ conductance feature that occurs in quantum point contacts. Our focus has been on the transconductance at finite source-drain bias which emphasizes transitions between the conductance plateaus. The calculated results based on  our model agree well with data taken on both ultra-low-disorder {\it induced} devices  and more traditional split-gated heterostructures. Without the inclusion of additional parameters this model also accounts for the observation of $0.7$ analogs  at high magnetic field. Although largely empirical, we believe the functional form underlying this simple picture may prove useful in uncovering a detailed microscopic theory of this effect.         

\section{Acknowledgments}
The authors wish to thank C. M. Marcus, T. M. Buehler, D. T. McClure, J. L. O'Brien, S. Das Sarma, K.-F. Berggren, M. J. Biercuk, R. G. Clark, A. Dzurak and A. R. Hamilton for useful discussions. D.J.R. is indebted to L. N. Pfeiffer and K. W. West for the excellent heterostructure material used in the experiments. 


\bibliographystyle{unsrt}

\end{document}